\documentclass[twoside,fleqn]{article}
\usepackage{npb}

\newcommand{\AmS}{{\protect\the\textfont2
  A\kern-.1667em\lower.5ex\hbox{M}\kern-.125emS}}

\newcommand{\GeV}{\mbox{ GeV}} 
 
\newcommand{\sq}{\ensuremath{\tilde q}} 

\newcommand{\stopp}{\ensuremath{\tilde t}} 
\newcommand{\cplus}{\ensuremath{\chi^+}} 
\newcommand{\neut}{\ensuremath{\chi^0}}

\newcommand{\sfr}{{\tilde{f}}}

\newcommand{\tb}{\ensuremath{\tan\beta}}

\newcommand{\mb}{\ensuremath{m_b}} 
\newcommand{\mt}{\ensuremath{m_t}} 
 
\newcommand{\mw}{\ensuremath{M_W}} 
\newcommand{\mws}{\ensuremath{M^2_W}} 
 
\newcommand{\mxs}{\ensuremath{M^2_X}}
\newcommand{\mx}{\ensuremath{M_X}}

\newcommand{\mHp}{\ensuremath{M_{H^\pm}}} 
 
\newcommand{\mi}{\ensuremath{M_i}} 
\newcommand{\mis}{\ensuremath{M^2_i}}

\newcommand{\ma}{\ensuremath{M^0_\alpha}} 
\newcommand{\mas}{\ensuremath{{M^0_\alpha}^2}} 

\newcommand{\mbet}{\ensuremath{M^0_\beta}} 
\newcommand{\mbets}{\ensuremath{{M^0_\beta}^2}} 

\newcommand{\mg}{\ensuremath{m_{\tilde{g}}}} 
 
\newcommand{\msbo}{\ensuremath{m_{\tilde{b}_1}}} 
\newcommand{\msbt}{\ensuremath{m_{\tilde{b}_2}}} 
 
\newcommand{\msto}{\ensuremath{m_{\tilde{t}_1}}} 
\newcommand{\mstt}{\ensuremath{m_{\tilde{t}_2}}}

\newcommand{\msdo}{\ensuremath{m_{\tilde d_1}}} %\alpha}}}
\newcommand{\msdt}{\ensuremath{m_{\tilde d_2}}} %\beta}}}
\newcommand{\msut}{\ensuremath{m_{\tilde u_2}}}

\newcommand{\mselo}{\ensuremath{m_{\tilde{e}_1}}} 
\newcommand{\mselt}{\ensuremath{m_{\tilde{e}_2}}} 
 
\newcommand{\mwf}{{\mw^4}}

\newcommand{\cbta}{c_\beta}
\newcommand{\cfbt}{c_{4\beta}}
\newcommand{\sbta}{s_\beta}
\newcommand{\stbt}{s_{2\beta}}

\newcommand{\sws}{s_W^2}

\newcommand{\UChaio}{U_{i1}}
\newcommand{\UChait}{U_{i2}}

\newcommand{\mchasomenysmchast}{\left(M_1^2-M_2^2\right)}
\newcommand{\commonnumfactor}{\frac{\alpha}{4\,\pi\,\sws}}

\newcommand{\msfos}{\ensuremath{m_{\tilde f_1}^2}}
\newcommand{\msfts}{\ensuremath{m_{\tilde f_2}^2}}

\newcommand{\osf}{\ensuremath{\theta_f}}
\newcommand{\osb}{\ensuremath{\theta_b}}
\newcommand{\ost}{\ensuremath{\theta_t}}
\newcommand{\osd}{\ensuremath{\theta_d}}
\newcommand{\osu}{\ensuremath{\theta_u}}
\newcommand{\osel}{\ensuremath{\theta_e}}

\newcommand{\SigmaSL}[1]{\ensuremath{\hat{\Sigma}_{SL}^{#1}}}
\newcommand{\SigmaSR}[1]{\ensuremath{\hat{\Sigma}_{SR}^{#1}}}
\newcommand{\SigmaL}[1]{\ensuremath{\hat{\Sigma}_{L}^{#1}}}
\newcommand{\SigmaR}[1]{\ensuremath{\hat{\Sigma}_{R}^{#1}}}

\title{Fermionic decays of sfermions in the MSSM: a full one-loop
  calculation
\thanks{Talk presented at \textit{6th International Symposium on
  Radiative Corrections Application of 
Quantum Field Theory to Phenomenology (RADCOR 2002) and 6th Zeuthen
Workshop on Elementary Particle Theory Loops and Legs in Quantum Field
Theory}, Kloster Banz, Germany, 8-13 September, 2002.}
}

\author{\underline{Jaume Guasch}\address{Theory Group LTP, Paul
    Scherrer Institut, CH-5232 Villigen PSI, Switzerland.},  
Wolfgang Hollik\address{Max-Planck-Institut f\"ur Physik,
   F\"ohringer Ring 6, D-80805 M\"unchen, Germany.},
 Joan Sol\`a\address{Departament d'Estructura i Constituents de la 
Mat\`eria,  Universitat de Barcelona, Diagonal 647, E-08028 Barcelona,
 Catalonia, Spain, 
and 
Institut de F\'{\i}sica d'Altes Energies, 
Universitat Aut\`onoma de
  Barcelona, E-08193 Bellaterra., Barcelona, Catalonia, Spain.}%
}

\begin{document}

\begin{abstract}
We present a full one-loop calculation of the electroweak  corrections to
the partial decay widths of the fermionic modes of sfermions
$\Gamma(\sfr \to f'\chi)$. The main technical points of the
renormalization are presented, and the main features of the results are
discussed. 
\begin{flushright}
{{PSI-PR-02-11, 
UB-ECM-PF-02-20,
%MPI-PhT/2002-??,
hep-ph/0210118
}}
\end{flushright}
\end{abstract}


\maketitle

\section{Introduction}
Supersymmetry (SUSY) is one of the firmest candidates for physics beyond the
Standard Model (SM) of particle interactions. If new particles are discovered
at the LHC, the forthcoming step will be to establish their nature. This
step could be performed at a  high energy $e^+e^-$ linear collider. One
of the basic predictions of SUSY is the equality of the couplings in the
different sectors: the SM gauge couplings must be equal to the
superpartners gauge couplings and to the Yukawa coupling between a
particle and its superpartner. 

We present a full one-loop computation of the electroweak (EW) corrections to
the sfermion partial decay width
\begin{equation}
\Gamma(\sfr \to f' \chi)\,\,,
\label{eq:gammadef}
\end{equation}
$\sfr$ being a sfermion, $f'$ a SM fermion and $\chi$ a chargino
($\chi^\pm_{1,2}$) or a neutralino ($\chi^0_{1\ldots4}$). The process~(\ref{eq:gammadef})
probes the fermion-sfermion-gaugino/higgsino coupling, and can be used
to test the SUSY relations. As we will see, the radiative corrections
induce finite shifts in these relations which are non-decoupling.

The QCD corrections to the process~(\ref{eq:gammadef}),                        
in the framework of the Minimal Supersymmetric Standard Model (MSSM), were     
computed                                                                       
in~\cite{QCD}. %\cite{Kraml:1996kz,Djouadi:1997wt}. 
However, the electroweak   
effects are much more cumbersome as their computation requires the             
renormalization of the whole MSSM Lagrangian. A first estimate was given       
in~\cite{Guasch:1998as}, where the contribution from the Yukawa couplings of   
bottom quarks decaying into chargino-neutralinos was analyzed within the       
so-called higgsino approximation. The electroweak effects have been further    
elaborated in~\cite{EWcorr}.                                                  
                                 
\section{Renormalization and radiative corrections}
The computation to one-loop level of the partial decay
width~(\ref{eq:gammadef}) requires the renormalization of the full MSSM
Lagrangian, taking into account the relations among the different
sectors and the mixing parameters. We choose to work in an on-shell
renormalization scheme, in which the renormalized parameters are the
measured quantities. 
The SM sector is renormalized according to the                                
standard on-shell SM $\alpha$-scheme~\cite{Hollik}, and the MSSM Higgs         
sector (in particular the renormalization of $\tb$) is treated as in~\cite{Dabels}.

As far as the sfermion sector is concerned, each squark doublet contains 5
independent parameters. We choose to use the following set of input
parameters\footnote{Throughout this work we make use of third generation
notation. The notation is as in~\cite{Guasch:1998as,EWcorr}.}: 
\begin{equation}
  \label{eq:inputsf}
(m_{\tilde{b}_1}, m_{\tilde{b}_2}, \osb ,m_{\tilde{t}_2},
\ost), \ \ \ \ m_{\sfr_1}>m_{\sfr_2}.  
\end{equation}
 The remaining parameters are computed as a function of those
 in~(\ref{eq:inputsf}). In particular, the trilinear soft-SUSY-breaking
 couplings read:
\begin{equation}
A_{\{b,t\}}=\mu\{\tan\beta,\cot\beta\}+
{m_{\tilde{f}_1}^2-m_{\tilde{f}_2}^2\over 2\,m_f}\,\sin{2\,\osf}\,,
\label{eq:Abt}
\end{equation}
with $\tb=v_2/v_1$, the ratio of the vacuum expectation values of the
two Higgs boson doublets. The mass parameters in~(\ref{eq:inputsf}) are
defined to be on-shell. For the mixing angle counterterm we adopt the
following convention:
\begin{equation}
\delta\osf=\frac{1}{2}\,\frac{\Sigma^{21}_{\tilde{f}}(\msfts)+\Sigma^{12}_{\tilde {f}}(\msfos)}{\msfts-\msfos} \,,
\label{eq:deltatheta}
\end{equation}
being $\Sigma^{ij}(k^2)$ the mixing self-energy between sfermions $i$
and $j$. The convention~(\ref{eq:deltatheta}) is known to introduce
gauge parameter dependence in the definition of the mixing
angle, 
but the result is the same as the gauge independent                        
computation of the self-energies                                               
in~(\ref{eq:deltatheta}) within the 't Hooft-Feynman                           
gauge~\cite{Yamada2}. 
Since the heaviest top-squark mass is not an input
parameter, it 
receives finite radiative corrections:
\begin{equation}
  \label{eq:mst1radcor}
  \Delta \msto^2 = \delta \msto^2 + \Sigma_{\stopp_1}(\msto^2) \,\,, %\neq 0 \,\,,
\end{equation}
where $\delta\msto^2$ is a combination of the counterterms of the
parameters in~(\ref{eq:inputsf}), and the counterterms of the gauge and
Higgs sectors. 

The chargino/neutralino sector contains six particles, but only three
independent input parameters ($M$ and $M'$ --the soft-SUSY-breaking
$SU(2)_L$ and $U(1)_Y$ gaugino masses-- and the higgsino mass parameter $\mu$). The situation in this sector is quite
different from the 
sfermion case,
since in this case no independent
counterterms for the mixing matrix elements can be introduced. We stick
to the following procedure: First, we introduce a set of
\textit{renormalized} parameters $(M,M',\mu)$ in the expression of the
chargino and neutralino matrices ($\cal{M}$ and ${\cal M}^0$), and
diagonalize them  by means of unitary matrices $M_D=U^* {\cal
  M} V^\dagger$, $M_D^0 = N^* {\cal M}^0 N^{\dagger}$. Now $U$, $V$ and
$N$ must be regarded as \textit{renormalized mixing matrices}. The
counterterm mass matrices are then $\delta M_D=U^* \delta{\cal M}
V^\dagger$, $\delta M_D^0 = N^*\delta {\cal M}^0 N^{\dagger}$, which are
non-diagonal. At this point, we introduce renormalization conditions for
certain elements of $\delta M_D$ and $\delta M_D^0$. In particular, using
the unitarity properties of $U$ and $V$ we can write the following set
of equations
\begin{equation}
\begin{array}{l}
M_1 \,\delta M_1 + M_2 \,\delta M_2 =  M \,\delta M + \mu \,\delta\mu+ \delta
\mws , \nonumber\\
 M_1 M_2
\left(M_1 \,\delta M_2+M_2\,\delta M_1\right)=\nonumber\\
\phantom{aaaaaa}=\left(M \mu -\mws
  \stbt\right) %\nonumber\\
\times\big[M \,\delta \mu + \mu \,\delta M \nonumber\\
\phantom{aaaaaaaaaa}-\mws \,\delta\stbt-\stbt \,\delta\mws\big] \,.
\end{array}
\label{eq:defcountcha} 
\end{equation}
$M_1$ and $M_2$ are the chargino on-shell masses, and $\delta M_i$
is a shortcut for the following combination of chargino self-energies:
\begin{equation}
\frac{  \delta
  M_i}{M_i}=-\frac{1}{2}\left(\Sigma_L^{-i}(M_i^2)+\Sigma_R^{-i}(M_i^2)\right)-\Sigma_S^{-i}(M_i^2)\,\,.
  \label{eq:defdeltami} 
\end{equation}
Conditions~(\ref{eq:defcountcha}) and~(\ref{eq:defdeltami}) are
equivalent to one-loop on-shell renormalization of the chargino masses.
As a last renormalization condition, we require on-shell renormalization
for one of the neutralinos, which we choose the lightest one, and find
the expression for the $M'$ counterterm:
\begin{equation}
  \label{eq:soldM1}
  \delta M'=\frac{1}{N_{11}^{*2}}(\delta M_1^0 -
  \sum_{\alpha\ {\rm or}\ \beta\neq1}N^*_{1 \alpha} \delta{\cal M}^0_{\alpha\beta} N_{
  1\beta}^{*} 
  )\,\,,
\end{equation}
where $\delta M_1^0$ is as in~(\ref{eq:defdeltami}), but for te lightest
neutralino ($\neut_1$).
Solving eqs.~(\ref{eq:defcountcha}) for $\delta M$ and $\delta\mu$, and
using eq.~(\ref{eq:soldM1}) for $\delta M'$, the renormalization
conditions for the chargino/neutralino sector are complete. The other
neutralino masses receive radiative corrections. In this framework the
renormalized one-loop 
chargino/neutralino 2-point functions 
are \textit{non}-diagonal. Therefore one must take into account this
mixing either by including explicitly the reducible $\chi_r-\chi_s$
mixing diagrams, or by means of external mixing wave-function terms,
e.g. the left- and right-chiral form factors for the creation of a neutralino
$\chi^0_\alpha$ must be multiplied by,
\begin{equation}
\begin{array}{l}
  \label{eq:neutmix}
  {\cal Z}_R^{0\beta \alpha}= \frac{1  }{\mas-\mbets} \times \\
~~~ [  \mbet\,\SigmaSL{\beta\alpha}(\mas) % \nonumber\\
    +\ma\,\SigmaSR{\beta\alpha}(\mas) \\%\nonumber\\&&
~~~    +\mbet\,\ma\,\SigmaL{\beta\alpha}(\mas) 
    +\mas\,\SigmaR{\beta\alpha}(\mas) ]  \,\,,%    \nonumber\\                 
%&&\nonumber\\
\\
   {\cal Z}_L^{0\beta \alpha}=\frac{1}{\mas-\mbets}\times \\
~~~ [ \mbet\,\SigmaSR{\beta\alpha}(\mas) 
     +\ma\,\SigmaSL{\beta\alpha}(\mas) \\
~~~     +\mbet\,\ma\,\SigmaR{\beta\alpha}(\mas)
     +\mas\,\SigmaL{\beta\alpha}(\mas) ]
   \,\,.
  \end{array}
\end{equation}
Here $\hat\Sigma^{\beta\alpha}_*(k^2)$ are the renormalized mixing
two-point functions  $\chi^0_\beta \to \chi^0_\alpha$. 
Equivalent expressions can be worked out for
the chargino sector. See Refs.~\cite{EberlFritzsche} %\cite{Eberl:2001eu,Fritzsche} 
for different (but
one-loop equivalent) approaches to the renormalization of the
chargino/neutralino sector.

The complete one-loop computation consists of:\\
%\begin{itemize}
%\item 
$\bullet$ renormalization constants for the parameters and wave functions in
  the bare Lagrangian,\\
%\item
$\bullet$ one-loop one-particle irreducible three-point functions, \\
%\item 
$\bullet$ mixing terms among the external charginos and neutralinos,\\
%\item 
$\bullet$ soft- and hard- photon  bremsstrahlung.\\
%\end{itemize}
All kind of MSSM particles are taken into account in the loops: SM fermions,
sfermions, electroweak gauge bosons, Higgs bosons, Goldstone bosons,
Fadeev-Popov ghosts, charginos, neutralinos. The computation is
performed in the 't Hooft-Feynman gauge, using dimensional reduction
for the regularization of divergent integrals. 
The loop computation itself is done using the computer algebra packages
\textit{FeynArts 3.0} and \textit{FormCalc
  2.2}~\cite{FeynArts3,Hahn:1998yk}. %\cite{Kublbeck:1990xc,Hahn:1998yk}. 
The numerical evaluation of one-loop integrals makes
use of \textit{LoopTools 1.2}~\cite{Hahn:1998yk}.\footnote{The resulting
  FORTRAN code can be obtained from \texttt{http://www-itp.physik.uni-karlsruhe.de/$\sim$guasch/progs/}.}

\section{Results}
The results show the very interesting property that none of the particles
of the MSSM decouples from the corrections to the
observables~(\ref{eq:gammadef}). This can be well understood in terms of
renormalization group (RG) running of the parameters and SUSY
breaking. Take, e.g., the effects of squarks in the
electron-selectron-photino coupling. Above the squark mass scale
($Q>m_{\sq}$) the electron electromagnetic coupling ($\alpha(Q)$) is equal
(by SUSY) to the electron-selectron-photino coupling
($\tilde{\alpha}(Q)$), and both couplings run according to the same RG
equations. At $Q=m_{\sq}$ the squarks \textit{decouple} from the
RG running of the couplings. At $Q<m_{\sq}$, $\alpha(Q)$ runs due to the
contributions from pure quark loops, but $\tilde{\alpha}(Q)$ does not
run anymore, and it is
\textit{frozen} at the squark scale, that is:
$\tilde{\alpha}(Q<m_{\sq})=\alpha(m_{\sq})$.  Therefore, when comparing
these two couplings at a scale $Q<m_{\sq}$, they differ by the 
logarithmic running of $\alpha(Q)$ from the squark scale to $Q$:
$\tilde{\alpha}(Q)/\alpha(Q)-1=\beta \log(m_{\sq}/Q)$.

The above discussion has two important consequences: \\
\textbf{1)} The
non-decoupling can be used to extract information of the high-energy
part of the SUSY spectrum: one can envisage a SUSY model in which a
significant splitting among the different SUSY masses exists,
e.g. $m_{\sq} \gg m_{\tilde l}$, where the sleptons lie below the
production threshold in an $e^+e^-$ linear collider, but the squarks are
above it. By means of high precision measurements of the
lepton-slepton-chargino/neutralino couplings one might be able to
extract information of the squark sector of the model, to be checked
with the available data from the LHC.\\
\textbf{2)} By the same token, it means
that the value of the radiative corrections depends on all
parameters of the model, and we can not make precise quantitative
statements unless the full SUSY spectrum is known. 
This drawback can be partially overcome by the introduction of                
\textit{effective coupling                                                     
matrices}, which can be defined as follows. The subset of fermion-sfermion     
one-loop contributions to                                                      
the self-energies of gauge-boson, Higgs-bosons, Goldstone-bosons, charginos    
and neutralinos form a gauge invariant finite subset of the corrections.       
Therefore these contributions can be absorbed into a finite shift of the       
chargino/neutralino mixing matrices $U$, $V$ and $N$
appearing in the
couplings: $
  U^{eff}=U+\Delta U^{(f)}   , \ 
  V^{eff}=V+\Delta V^{(f)}    ,  \   N^{eff}=N+\Delta N^{(f)}$. In this
  way we can \textit{decouple} the computation of the \textit{universal}
  (or \textit{super-oblique}~\cite{Katz:1998br}) corrections. These
  corrections contain the 
  non-decoupling logarithms from sfermion masses. We have been able to
  derive an analytic expression containing these logarithms in a simple
  case. We have computed  the electron-selectron contributions to
  $\Delta U^{(f)}$  and $\Delta V^{(f)}$
matrices, assuming zero mixing angle in the
selectron sector ($\theta_e=0$), we have identified the leading terms in the
approximation $m_{\tilde e_i}, m_{\tilde \nu}\gg (\mw,\mi) \gg m_e$, and
 analytically cancelled the divergences and the renormalization scale
 dependent terms;
finally, we have kept only the terms logarithmic in the {slepton}
masses. The result for $\Delta U^{(f)}$ reads as follows:
\begin{eqnarray}
\begin{array}{l}
\Delta\UChaio^{(f)}= \commonnumfactor\log\left(\frac{M^2_{\tilde e_L}}{\mxs}\right)\,\bigg[
 \frac{\UChaio^3}{6} - 
\\ 
 ~~~~ \UChait \frac{\sqrt{2}\,\mw\,(M\,\cbta +
  \mu\,\sbta)}{3\,(M^2-\mu^2)\,\mchasomenysmchast^2} %\times \nonumber\\&&
\left(M^4 - M^2\,\mu^2 + \right.\nonumber\\
~~~\left. + 3\,M^2\,\mws + 
     \mu^2\,\mws + \mwf + \mwf\,\cfbt %\right.\nonumber\\&&
 \right.\\\left.
~~~   + (\mu^2-M^2)\,\mis + 
     4\,M\,\mu\,\mws\,\stbt\right)\,\bigg]\,\,,
 \end{array}
 \nonumber\\
\begin{array}{l}
\Delta\UChait^{(f)}=\commonnumfactor\log\left(\frac{M^2_{\tilde e_L}}{\mxs}\right)\,
\UChaio 
\times \\ ~~~\times
\,\frac{\mw\,(M\,\cbta + \mu\,\sbta)}
 {3\,\sqrt{2}\,(M^2-\mu^2)\,\mchasomenysmchast^2} 
%\times \nonumber\\
 % \times 
\left((M^2-\mu^2)^2 
+ \right. \\\left.
 ~~~+4\,M^2\,\mws + 4\,\mu^2\,\mws + 2\,\mwf + 
 \right.\\\left.~~~+  2\,\mwf\,\cfbt + 8\,M\,\mu\,\mws\,\stbt\right)\,\,,
\end{array}
\label{eq:logterms}
\end{eqnarray}
$M^2_{\tilde e_L}$ being the soft-SUSY-breaking mass of the
$(\tilde{e}_L,\tilde{\nu})$ doublet,
whereas $\mx$ is a SM mass.
In the on-shell scheme for the SM electroweak theory we define
parameters at very different scales, basically $\mx=\mw$ and
$\mx=m_e$.  These wide-ranging scales enter the structure of the
counterterms and so
must appear in eq.(\ref{eq:logterms}) too. As a result the leading log
in the various terms of this equation will vary accordingly. For
simplicity in the notation we have factorized $\log M^2_{\tilde
e_L}/\mxs$ as an overall factor. In some cases this factor can be very
big, $\log M^2_{\tilde e_L}/m_e^2$; it comes from the electron-selectron
contribution to the chargino-neutralino self-energies.

Although above we have singled out the non-decoupling properties of
sfermions, we would like to stress that the whole spectrum shows
non-decoupling properties. By numerical analysis we have been able to
show the existence of logarithms of the gaugino mass parameters ($M/M_X$ and
$M'/M_X$), and the Higgs mass ($\mHp/M_X$). However, due to the
complicated mixing 
structure of the model, we were not able to derive simple analytic
expressions containing these non-decoupling logarithms. 
Note that in \textit{any} observable which includes the
fermion-sfermion-chargino/neutralino Yukawa couplings at leading order
we will have this 
kind of corrections, therefore the full MSSM spectrum must be taken into
account when computing radiative corrections, since otherwise one could
be missing large logarithmic contributions of the heavy masses.

\begin{table}
  \centering
  \begin{tabular}{|c|c|c|c|c||c|}
%\hline
%\multicolumn{6}{|c|}{$BR(\tilde{b} \to q\chi)$}\\
\hline  
 & $\neut_1$ & $\neut_2$ & $\neut_3$ & $\neut_4$ & $\cplus_1$ %&
                                %$\cplus_2$
\\ \hline  
\multicolumn{6}{|c|}{$BR(\tilde{b}_1 \to q\chi)$}\\\hline
Tree %$BR^0(\sbottom_1)$ %$BR(\tilde{b}_1 \to q\chi)$  
& 0.272 & 0.092 & 0.047 & 0.014 & 0.575 %& -
\\ \hline  
QCD %$BR^{QCD}(\sbottom_1)$ %(\tilde{b}_1\to q\chi)$ 
& 0.308 & 0.104 & 0.031 & 0.018 & 0.538
                                %& - 
\\ \hline 
Total %$BR^{total}(\sbottom_1)$%(\tilde{b}_1\to q\chi)$ 
& 0.291 & 0.092 & 0.031 & 0.018 &
0.568 %& -
\\ \hline \hline 
\multicolumn{6}{|c|}{$BR(\tilde{b}_2 \to q\chi)$}\\\hline
Tree %$BR^0(\tilde{b}_2)$% \to q\chi)$
 & 0.502 & 0.332 & 0.123 & - & 0.042 %& -
\\
 \hline  
QCD %$BR^{QCD}(\tilde{b}_2)$ %\to q\chi)$ 
& 0.541 & 0.386 & 0.054 & - & 0.019 % & -
\\ \hline 
Total %$BR^{total}(\tilde{b}_2)$ %\to q\chi)$ 
& 0.528 & 0.395 & 0.056 & - & 0.020 %&-
\\ \hline 

  \end{tabular}
  \caption{Branching ratios of bottom-squarks into charginos and
    neutralinos for the parameter set~(\ref{eq:inputpars}). Shown are:
    the tree-level value, the QCD-corrected 
    value, and the value including EW corrections.
     Branching ratios below $10^{-3}$ are not shown.}
  \label{tab:corr}
\end{table}

As for the \textit{non-universal} part of the contributions, we summarize
here the main features: \\
\textbf{a)} The corrections grow as the logarithm squared of the decaying
particle mass, due 
to the presence of electroweak Sudakov double-logs~\cite{sudakov}.
\\
\textbf{b)} The corrections show multiple threshold singularities when
varying any of the parameters of the model, however they are well behaved
in the regions between thresholds. Therefore, in order to give a
quantitative value of the corrections the correlation between the
different particle masses must be known.
\\ 
\textbf{c)} The third  sfermion family contains a large contribution that can be
traced back to the presence of corrections similar to the 
\textit{threshold-like} corrections to the bottom-quark (and
$\tau$-lepton) Yukawa coupling $\Delta m_{\{b,\tau\}}$~\cite{DmbTeo}. This
contribution contains (aside from $\Delta m_{\{b,\tau\}}$)
terms from the sfermion mixing self-energies $\sfr_L - \sfr_R$,
and is (approximately)
proportional to the sfermion mass splitting. It is also present in the
QCD corrections.\\ 
\textbf{d)} The three-point vertex functions contain the
soft SUSY-breaking trilinear Higgs-sfermion-sfermion couplings $A_f$. If
$A_f$ is large, it would induce large corrections (even larger than
$100\%$!). However, as long as 
$A_f$ is in the perturbative coupling regime, the corrections stay below
$20\%$. \\
Points {c)} and {d)} play a complementary role in the
large $\tb$ regime due to the relation~(\ref{eq:Abt}). A small mass
splitting (or small mixing angle) means small $\Delta m_{\{b,\tau\}}$
corrections, but a large value of $A_{\{b,\tau\}}\simeq \mu\tb$. However
a large $A_f$ value would both spoil perturbativity and generate charge
and/or colour breaking vacua. Therefore, at large $\tb$ the
bottom-squarks and $\tau$-sleptons must have significant splitting and
mixing angle, providing large corrections of the type {c)}
above. 

The $\Delta m_{\{b,\tau\}}$ corrections  play also an
important role in the phenomenology of the MSSM Higgs bosons at large
$\tb$, showing non-decoupling properties for $M_{SUSY}\gg
M_{SM},M_{H^\pm}$, as stressed e.g. in~\cite{SUSYtbH}. The
$\sfr_a-f'-\chi_r$ coupling, on the other hand, exhibits a new kind of
non-decoupling properties, which are independent of the value of $\tb$.

As an example of the numerical value of the corrections, we present in
Table~\ref{tab:corr} the tree-level and corrected branching ratios for
bottom-squarks, for the following set of input parameters:
\begin{equation}
\begin{array}{l}
 \mt=175\GeV\,\,,
 \mb=5\GeV\,\,,
 \tb      =  4\,\,,\\
 \msbt=\msdt=\msut=\mselt=300\GeV\,\,,\\
 \msbo=\msdo=\mselo=\msbt+5\GeV\,\,,
\\\msut=290\GeV\,\,,
 \mstt=300\GeV\,\,,\\
 \osb=\osd=\osu=\osel=0\,\,,
 \ost=-\pi/5\,\,,\\
 \mu      =  150\GeV\,\,,
 M       =  250\GeV\,\,,\\
 \mHp  =  120\GeV\,\,,\mg=500\GeV\,\,.\\
\end{array}
\label{eq:inputpars}
\end{equation}
The results of Table~\ref{tab:corr} show that a high precision
measurement of the partial decay branching ratios of sfermions, to be
performed at a high energy $e^+ e^-$ linear collider, is sensitive to
the EW corrections.
From Table~\ref{tab:corr} it is also clear that the EW corrections can induce
a change on the branching ratios of the leading decay channels
comparable to the QCD corrections. Therefore both contributions must be
taken into account on equal footing in the analysis of the phenomenology
of sfermions.

\vspace{.2cm}

\noindent{Acknowledgments:}
The calculations have been done using the QCM cluster of the 
DFG Forschergruppe ``Quantenfeldtheorie, Computeralgebra und
Monte-Carlo Simulation''.
This collaboration is part of the network ``Physics at Colliders'' of the
European Union under contract HPRN-CT-2000-00149.
The work of J.G. has been partially supported by the 
European Union under contract No. HPMF-CT-1999-00150. 
The work of J.S.
has been supported in part by MECYT and FEDER under project FPA2001-3598.

\providecommand{\href}[2]{#2}

\end{document}